\documentclass{aa}  
\usepackage{graphicx}
\usepackage{lipsum}
\usepackage{txfonts}
\usepackage{comment}
\usepackage{subfigure}
\usepackage{xcolor}
\usepackage{afterpage}
\usepackage{siunitx}
\usepackage{academicons}
\usepackage{float} 
\usepackage{changepage}
\usepackage{lmodern}

\usepackage{setspace} 
\usepackage[breaklinks,colorlinks=true,linkcolor=blue,citecolor=blue]{hyperref}

\begin{document} 

   \title{Electron Beam Propagation and Radio-Wave Scattering in the Inner Heliosphere using Five Spacecraft}
   \titlerunning{Electron Beam Propagation and Radio-Wave Scattering in the Heliosphere}

   \author{L.~A.~Ca\~nizares \orcid{0000-0003-4711-522X}
          \inst{1,2} 
          \and
          S.~T.~Badman\inst{3} \orcid{0000-0002-6145-436X}
          \and
          N.~Chrysaphi\inst{4,5} \orcid{0000-0002-4389-5540}
          \and
          S.~Bhunia\inst{1,2} \orcid{0000-0001-8496-252X}
          \and
          B.~Sánchez-Cano\inst{6} \orcid{0000-0003-0277-3253}
          \and
          S.~A.~Maloney\inst{1} \orcid{0000-0002-4715-1805}
          \and
          P.~T.~Gallagher\inst{1} \orcid{0000-0001-9745-0400}
          }

   \institute{Astronomy \& Astrophysics Section, DIAS Dunsink Observatory, Dublin Institute for Advanced Studies, Dublin D15XR2R, Ireland.
         \and
             School of Physics, Trinity College Dublin, Dublin 2, Ireland.
         \and 
             Center for Astrophysics, Harvard \& Smithsonian, Cambridge, Massachusetts, US.
         \and 
             Sorbonne Universit\'{e}, \'{E}cole Polytechnique, Institut Polytechnique de Paris, CNRS, Laboratoire de Physique des Plasmas (LPP), 4~Place Jussieu, 75005 Paris, France.
         \and
            School of Physics \& Astronomy, University of Glasgow, Glasgow, G12 8QQ, UK.
         \and 
            University of Leicester, School of Physics and Astronomy, University Road, LE1 7RH, Leicester, UK.
             }

   \date{Received Month DD, YYYY; accepted Month DD, YYYY}
    
  \abstract
   {Solar energetic particles such as electrons can be accelerated to mildly-relativistic velocities in the solar corona. These electrons travel through the turbulent corona generating radio waves, which are then severely affected by scattering.}
   {The physical interpretation of the discrepancies between the actual and observed radio sources is still subject to debate. Here, we use radio emission observed by an unprecedented total of five spacecraft, to track the path of radio sources from the low corona to the inner heliosphere (15--75~R$_\odot$/ 0.07--0.35~au) generated during a solar event on 4 December 2021.} 
   {In this study we make use of the Bayesian multilateration technique known as BELLA to track the apparent path of radio sources observed by Parker Solar Probe, STEREO A, Wind, Solar Orbiter and Mars Express. To validate the accuracy of the tracked path, we used Nançay Radioheliograph interferometric imaging at 150~MHz, which was found to agree with the estimated footpoints predicted by BELLA. We also further validated our results using ACE in-situ measurements.}
   {We found that the apparent radio sources followed the path of an Archimedean Parker Spiral, with an associated solar wind velocity of approximately 493 km s$^{-1}$ (consistent with the corresponding speed observed at 1~au at the relevant longitude) and connected to 75$^\circ$ longitude East at the solar surface. Finally, we made quantitative estimates of the scattering of radio waves that were found to be in good agreement with contemporary models of scattering in which the radio waves primarily propagate along the local Parker spiral.} 
   {This work shows conclusive evidence that the disputed cause of the widely observed `higher than expected' electron densities at interplanetary distances is due to radio wave scattering, and provides a more detailed understanding of the propagation of radio waves emitted near the local plasma frequency in turbulent astrophysical plasmas.}

   \keywords{bayesian --
                triangulation --
                solar radio burst -- 
                solar corona --
                multilateration
               }

    \maketitle
     
    %
\nolinenumbers

\section{Introduction}\label{sec:intro}
Solar radio bursts (SRBs) occur as the result of mildly relativistic particles travelling through plasma in the solar corona. These SRBs serve as a probe to study the different acceleration and propagation mechanisms of such particles. The most prolific bursts are the Type~III, associated with electrons travelling along open magnetic field lines \citep[e.g.,][]{alvarez1972evidence, sw:reid2020review,sw:wang2023solar}, and the acceleration mechanisms of their drivers are often associated with flares \citep{sw:reid2017coronal}, although this is not always the case \citep{tri:badman2022tracking}.   

Type~IIIs are particularly interesting because they are observed at a wide range of frequencies with ground-based telescopes such as the Low Frequency Array \citep[LOFAR,][]{inst:vanHaarlem2013lofar} (240 -- 10~MHz) or Nançay Radio Heliograph \citep[NRH,][]{inst:kerdraon2007nanccay} (450 -- 150~MHz) which observe Type~IIIs in the low corona at metric to decametric wavelengths, and space-based instruments such as Solar Orbiter's Radio and Plasma Waves \citep[SolO/RPW,][]{inst:maksimovic2020rpw} and Parker Solar Probe's FIELDS \citep[PSP/FIELDS,][]{inst:Bale2016Fields} which observe interplanetary Type~IIIs in the decametric to kilometric wavelength range (20 -- 0.01~MHz).  Tracking Type III sources is relatively common at either the low corona  \citep[e.g.,][]{sw:morosan2014lofar, sw:mann2018tracking} or at the interplanetary range \citep[e.g.,][]{tri:cecconi2007influence, bay:Canizares2024}. However, with a few exceptions  \citep[e.g.,][]{tri:badman2022tracking} it is relatively uncommon to find studies that combine both ground and space-based observations to track the sources of a single Type III, from the low corona to interplanetary distances.  One reason for the lack of examples of sources being tracked from low corona to interplanetary distances is that the fleet of solar-dedicated spacecraft is limited, and as it will be shown in Sec.~\ref{sec:propagation}, there is a lack of understanding of the effects of spacecraft configuration on the localisation of the Type~III sources. This can confer significant uncertainties resulting in discrepancies between ground and space-based localisations. In addition, a lack of accurate and precise localisations makes the measurement of radio scattering a challenging task, further contributing to the unexplained discrepancies between the two.  

There are multiple methods used to localise the sources of Type~IIIs. Ground based telescopes such as LOFAR and NRH have the advantage of possessing large facilities, extensive instrument networks, and vast data processing capabilities. This results in intricate telescopes with subsecond to second cadence specifications \citep{inst:murphy2021first} capable of performing interferometric imaging with uncertainties of merely a few arcseconds. However, frequencies under $\sim$10~MHz are absorbed due to the Earth's ionospheric cut--off, making space-based telescopes essential for localising sources at interplanetary distances. Unfortunately, the current fleet of solar spacecraft is not capable of performing interferometric imaging and therefore other techniques are required.

Goniopolarimetry \citep[GP;][]{tri:manning1980new, tri:cecconi2008stereo} is a well-established method of localisation in which the polarisation of radio waves can be analysed to obtain the direction $\vec{k}$ of the radio waves and back propagating the vectors from multiple spacecraft results in the radiation source location \citep[e.g.,][]{tri:krupar2012goniopolarimetric}. The advantage of this method is that the minimum number of spacecraft required to obtain a solution is two. However, not all radio spectrometers are designed with GP capabilities and the method is relatively susceptible to antenna anomalies \citep{Bonnin_Antenna3_Anomaly_Report_2024}. An alternative method is multilateration \citep{tri:Alcock2018, tri:badman2022tracking} which requires an additional spacecraft but benefits from using time information only. This makes it compatible with all radio spectrograms and makes the data processing relatively simple. There are two main types of multilateration, Time-of-Arrival~(TOA) and Time-Difference-of-Arrival~(TDOA). TOA consists of subtracting the arrival time from the emission time to derive multiple circles in which the solution is found at the intersection of these. On the other hand, TDOA does not need the emission time, which is intrinsically unknown in the case of SRBs, and uses the time difference between the arrival times to derive a hyperbolic geometrical function for every spacecraft pair where the solution is obtained at the intersection of three or more hyperbolic functions. The biggest disadvantage of all these methods is that computing the uncertainties is not a straightforward task and therefore statistical methods such as the BayEsian LocaLisation Algorithm \citep[BELLA,][]{bay:Canizares2024} need to be employed (see Sec.~\ref{sec:methods:bella}). 

Previous studies have shown \citep[e.g.,][]{tri:chen2023source, bay:Canizares2024} that the apparent location of Type~III sources is typically located much further from the Sun than expected by empirical density models. \cite{scat:chrysaphi2018cme} and \cite{scat:kontar2019anisotropic} have quantitatively shown that this is expected as a consequence of radio scattering. Type~III SRBs produce waves via the plasma emission mechanism \citep[see reviews for more details,][]{sw:melrose1987plasma,sw:reid2014review}, which produces radio emission at near the plasma frequency. Given that the refractive index of an unmagnetised plasma is $n = (1- \omega_{pe}^2/\omega^2)^{1/2}$ \citep[see ][]{scat:kontar2019anisotropic}, the emission of waves produced at near the plasma frequency $\omega \approx \omega_{pe}$ will result in a refractive index $0 \leq n < 1$. These micro-refractions produce a macroscopic radio scattering effect that was quantitatively estimated by \cite{scat:chrysaphi2018cme} using an analytical solution based on optical depth. These studies also showed that the effects of scattering are inversely correlated with radial distance, and demonstrated that at low frequencies (and even at high frequencies), these effects are too large to be ignored. 

In this paper, we present a Type~III radio burst and track it from low in the corona ($\sim$~0.25~R$_\odot$) using NRH interferometric imaging at 150~MHz to interplanetary distances using BELLA at 3--0.5~MHz with a localisation precision at the interplanetary range of $\sim$~15--30~R$_\odot$, following the path of a Parker Spiral with a fit quality of 99.2\%. In Sec.~\ref{sec:methods} we briefly introduce the different methods used in this study. In Sec.~\ref{sec:observations}, we will explore the wide range of observations needed to track the path of these electron beams using a five spacecraft observation of a Type~III SRB, thanks to the addition of Mars Express' Mars Advanced Radar for Subsurface and Ionosphere Sounding \citep[MEX/MARSIS,][]{inst:jordan2009marsis}. Sec.~\ref{sec:accel} will explore the possible acceleration mechanism that triggered the electron beams, and in Sec.~\ref{sec:propagation} we will show the propagation path of the Type~III exciters, demonstrating the important implications of understanding spacecraft configuration. Finally, in Sec.~\ref{sec:scattering} we will utilise these localisation measurements to make quantitative estimations of the radio wave scattering, showing strong agreement with predicted theoretical values. 

\section{Methods}\label{sec:methods}
\subsection{BELLA}\label{sec:methods:bella}
\begin{figure*}[h]
    \centering
    \includegraphics[width=1\linewidth]{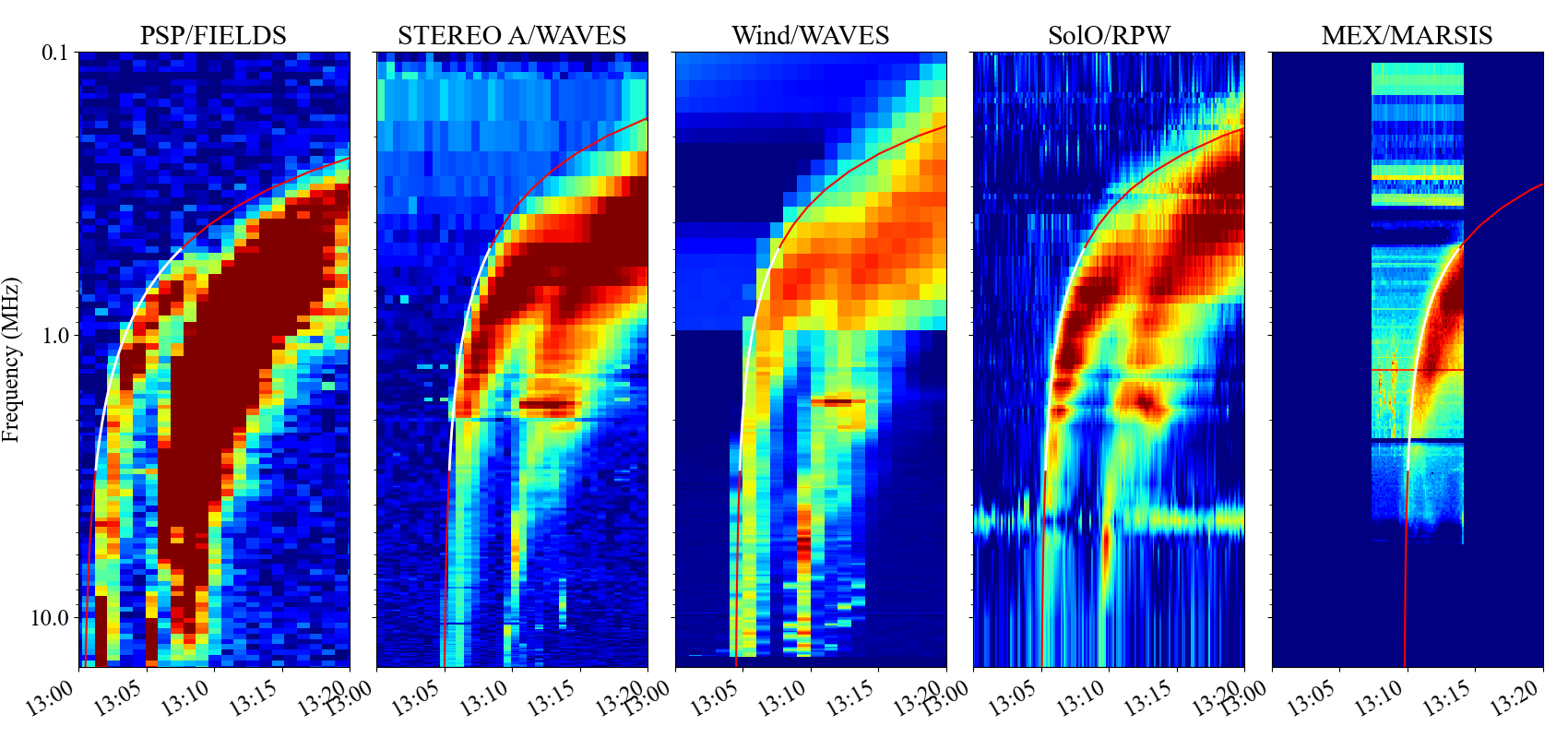}
    \caption{Double Type~III radio burst observed, from left to right, by PSP/Fields, STEREO A/Waves, Wind/Waves, SolO/RPW, and MEX/MARSIS on 4 December 2021. MEX/MARSIS only observed the earlier burst due to the narrow window of operation of MARSIS. Because of this limitation, only the earlier burst was considered for this study. The front of the burst in question was obtained by fitting an inverse polynomial function to manually extracted data (red line). Due to the frequency range limitations of MARSIS, the extracted fit for the Bayesian multilateration was kept at a conservative 0.5--3~MHz (white segment).}
    \label{fig:dynspec}
\end{figure*}

The Type~III SRBs from Fig.~\ref{fig:dynspec} were processed with BELLA. The automatic detection method used by \cite{bay:Canizares2024} to obtain the leading edge of the bursts was not utilised in this case because there are two Type~IIIs merging at the low frequencies and the automatic detection method is not designed for spectra with more than one burst. The data was obtained by manually recording ten data points from the dynamic spectra and then fitting an inverse polynomial to the data to characterise the front of the Type~III \citep{bay:Canizares2024}:
\begin{equation}\label{eq:timeevo}
    t(f) = a_2\frac{1}{f^2} + a_1\frac{1}{f} + a_0,
\end{equation}
where t is the timestamp, f is the frequency and ($a_2$, $a_1$, $a_0$) are the fitting parameters that describe the drift rate, curvature and starting point of the SRB \citep{bay:Canizares2024}. 

Once each Type~III was characterised, 50 timestamps were obtained using this front fitting function in a range between 0.5~MHz and 3~MHz. 

The BELLA-Multilaterate pipeline~\citep{soft:luis_alberto_canizares_alberto_2023_10276815} was employed once the timing information was obtained. The default BELLA parameters were used to achieve convergence. The default BELLA cadence is of 60~s, with 4 chains of 2000 tuning samples and 2000 standard samples. BELLA utilises the PyMC python package to perform Bayesian multilateration which according to \cite{bay:Canizares2024} is defined as:
\begin{equation}\label{eq:bayesianEQ}
    P(\vec{x},v \ | \ \Delta t) =  \frac{P(\Delta t \ | \ \vec{x},v) \  P(\vec{x} \ | \ v ) \ P(v)}{ P(\Delta t)},
\end{equation}
where each term is defined in \cite{bay:Canizares2024} and is summarised as:
\begin{itemize}
    \item $P(\vec{x},v \ | \ \Delta t)$ the posterior probability (i.e. the location of the source)
    \item $P(\Delta t \ | \ \vec{x},v)$ the likelihood function based on the physics model employed (i.e. $t~=~d/v$)
    \item $P(\vec{x} \ | \ v ) $, the source position prior distribution. 
    \item $P( v ) $, the propagation velocity prior. 
    \item $P(\Delta t)$, the observable time information. 
\end{itemize}
The default BELLA priors were employed and are available at \cite{bay:Canizares2024}. 

The background confidence maps were obtained by defining a grid of [-310, 310]~R$_\odot$ in both the x and y direction to account for the position of MEX. The resolution of the grid for both the x and y coordinates was set at 10~R$_\odot$. The cadence of the simulations was set at 60~s because it is the cadence of the worst available spectrogram.

\subsection{NRH Interferometric Imaging}
The NRH images were processed using the SolarSoftWare \citep[SSW,][]{other:freeland1998data} NRH IDL software. NRH has the capability of imaging at several frequencies (444.0, 432.0, 408.0, 370.0, 327.0, 298.7, 270.6, 228.0, 173.2, and 150.9 MHz) at a time resolution of 0.25 s. 128$\times$128 pixels (each pixel size $\approx15.17^{\prime\prime}$) of 2D intensity images were produced at 150.9~MHz. The imaged sources were consistently located at approximately 0.25~R$_\odot$ for a total time of 9~s. 

\subsection{PFSS}
The PFSS was computed with the help of pfsspy\footnote{Now housed under \texttt{sunkit-magex} at \url{https://github.com/sunpy/sunkit-magex}} \citep{other:Stansby2020} using GONG data obtained via SunPy's Fido routine. Given that the flaring event occurred on the western limb, a 90\% mask was applied to the PFSS for the purposes of showing field lines from the limb only. For display purposes, the closed field lines were filtered out.

\subsection{The Parker density model}
The Parker density model is a theoretical model described in \cite{dm:parker1960} that is appropriate at interplanetary distances where the magnetic field configuration is assumed to be a spiral in shape. This model was modified to match observations made by \cite{dm:mann1999} and is described in \cite{scat:kontar2019anisotropic} as,
\begin{equation}\label{eq:parkermodel}
    n(r) = 4.8 \times 10^{9} \left(\frac{R_\odot}{r}\right)^{14} + 3 \times 10^{8} \left(\frac{R_\odot}{r}\right)^{6} + 1.4 \times 10^{6} \left(\frac{R_\odot}{r}\right)^{2.3}
\end{equation}

\subsection{Radio-wave propagation simulations}\label{sec:methods:scatteringsimulations}
This study makes use of 3D ray-tracing simulations of radio-wave propagation in a medium of anisotropic density fluctuations, as presented in \cite{scat:kontar2019anisotropic}. 
We adopted the upgraded version of the simulations (available as open source) that allow for a description of a magnetic field that follows the Parker Spiral (as done in \cite{tri:chen2023source} and \cite{scat:chrysaphi2024first}).
A solar wind speed of 420~km/s at 1~au was assumed in the simulation of the Parker Spiral, and both fundamental ($f=1.1f_{pe}$) and harmonic ($f=2f_{pe}$) emissions were simulated using 1e5 photons. 
The strength of scattering was defined as $\overline{q\varepsilon^2} = 1$ and the level of anisotropy was assumed to be $\alpha=0.25$ for fundamental emissions and $\alpha=0.4$ for harmonic emissions, in line with the average values found to match a variety of observations across a large range of frequencies, including solar and extra-solar radio measurements and in-situ density measurements (see \cite{scat:kontar2023anisotropic} for details).

\section{Multi-instrument observations}\label{sec:observations}
On 4 December 2021 at $\sim$~13:00~UT two consecutive Type~IIIs were detected by PSP/FIELDS, Solar TErrestrial RElations Observatory A~\citep[STEREO/Waves,][]{inst:Bougeret2008swaves}, Wind~\citep[Wind/Waves,][]{inst:Bougeret1995waves}, SolO/RPW and MEX/MARSIS, see Fig.~\ref{fig:dynspec}. This is the first time that Type~III SRBs have been reported from Mars Express. In particular, MARSIS \citep{inst:gurnett2005radar} is able to detect them in its Active Ionospheric Sounding mode between 0.1 and 5.5 MHz. A study is currently ongoing to catalogue all these events. The Type~IIIs, were separated by $\sim$~5~min and there was no evidence of fundamental-harmonic ratios in their morphology implying that they were two separate events and not a fundamental-harmonic pair. This was further confirmed by observing the eCallisto \citep{inst:benz2004callisto} spectrogram shown in Fig.~\ref{app:ecallisto} of the appendix.

MARSIS comes with a number of limitations that restrict the utilisation of this instrument for solar studies. Firstly, MARSIS is only operational for up to 40 minutes at a time and as shown by Fig.~\ref{fig:dynspec}, the window of operation was reduced to $\sim$~10~min for this particular observation. Unfortunately, the second burst was missed by MARSIS and therefore this study will focus on the first burst which was observed by all five spacecraft. In order to confirm that the MARSIS detected burst was indeed the earlier burst, we light-travel corrected all spectra to 1~au and stacked the spectra vertically. The second limitation of MARSIS is the narrow frequency range compared to that of the solar dedicated missions. Furthermore, in addition to MARSIS' narrow frequency range, Fig.~\ref{fig:dynspec} shows that the burst is weakly detected at frequencies larger than 3~MHz. This could be attributed to a number of effects, such as the larger distance of Mars compared to other observatories, and also, could be a consequence of instrumental sensitivity due to the orientation of the antenna. Given its dipolar nature, MARSIS is least sensitive along the axis of the antenna and most sensitive in the perpendicular direction. Therefore, the orientation of MEX relative to the sources, could affect the SRB recording. Additionally, because of the narrow operational window of MARSIS, the data appears to be truncated at the lower frequencies therefore setting a lower frequency limitation at $\sim$~0.5~MHz. 

 Looking at the solar dedicated spectrograms, the lower frequency limitation imposed by MARSIS was inevitable as FIELDS also did not observe the earlier burst at lower than 0.5~MHz, and the other spectrograms showed that the later burst merged with the earlier burst at $\sim$~0.3~MHz. Despite these limitations, MARSIS provided spectrograms with a cadence of 7.5~s, making it the highest time resolution spectrogram of the five, followed by RPW at 7.6~s and 9.2~s for HFR/TNR respectively, SWAVES at 34.98~s, FIELDS was set up at 55.92~s for this observation, and the WAVES available data had a cadence of 60~s. 

To perform BELLA's Bayesian Multilateration \citep[see][for details]{bay:Canizares2024} time information from the same frequency channel is required. However, all the spectrograms contain different frequency channels. In order to mitigate this mismatch in frequency channels, an inverse polynomial function is fitted to the leading edge of the Type~III SRBs (see Eq.~\ref{eq:timeevo}). This front-fitting function is fitted from manually selected data and is plotted in Fig.~\ref{fig:dynspec} as a red line. The segment of the red-fitted function used for the multilateration is shown as white. This white segment corresponds to the overlapping frequency range of all the spectrograms. 

 The data obtained from the Geostationary Operational Environmental Satellite \citep[GOES,][]{sc:mallette1982geostationary} and The Spectrometer Telescope for Imaging X-rays~\citep[SolO/STIX,][]{inst:krucker2020spectrometer_solo_stix} (available in the Appendix as Fig.~\ref{app:xray}), showed a small B class flare detected by GOES at 1.0--8.0~\AA, and by STIX in the 4--10~keV regime. A red vertical line shows the detection time of the Type~III SRB by NRH which is consistent with the onset of the flaring event detected by the X-Ray spectrometers.  

At approximately the same time as the Type III emission observed in Fig.~\ref{fig:dynspec}, the NRH detected a source in the 150 MHz frequency channel. Additionally, Fig.~\ref{app:ecallisto} in the appendix shows RPW data together with e-CALLISTO spectra from Birr, Glasgow, and Humain \citep{inst:benz2004callisto}. Emission from both Type IIIs was detected in the range of 20--80 MHz. Unfortunately, radio frequency interference obscured the 100-150 MHz frequency domain. However, on that day, the only active region producing flares was AR 12898 \footnote{\url{https://solmon.dias.ie/fulldisk?date=2021120413}}. Furthermore, Fig.~\ref{app:ecallisto}b shows that the apparent movement of the sources shown in Fig.~\ref{fig:dynspec} are likely attributed to multiple Type IIIs. Ionospheric disturbances could potentially cause these shifts if the time scale variations are in the order of minutes \citep{inst:mercier1986observations, sw:jordan2017characterization_ionos}, however, the imaged sources are all occurring within seconds of each other, and it is therefore unlikely.

\begin{figure}[h!]
    \centering
    \includegraphics[width=1\linewidth]{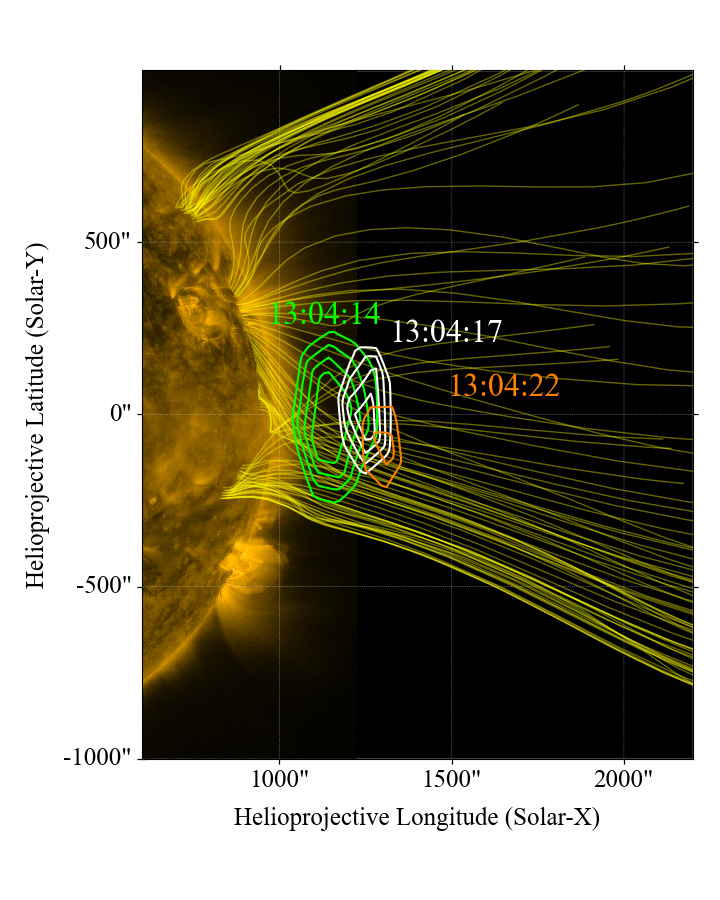}
    \caption{NRH imaging of the Type III radio burst sources at 150~MHz, overlaid on an AIA 171~\AA \, image, together with a PFSS derived from GONG. The 150~MHz sources, consistently observed at approximately $\sim$~0.25~$R_\odot$, align with the multilateration results. The radio emission is associated with a B-class flare, as shown in Fig.~\ref{app:xray}. Additionally, the PFSS shows a large group of open magnetic field lines at the 150 MHz location consistent with the Parker Spiral derived from the multilateration results. }
    \label{fig:imaging}
\end{figure}

\section{Results and discussion}\label{sec:results}
\subsection{Onset of the electron beams}\label{sec:accel}

Fig.~\ref{fig:imaging} shows NRH imaging at 150~MHz overlaid on a Potential-Field Source-Surface (PFSS)\citep{other:Stansby2020} and image taken by AIA at 171~\AA \, at 13:04:14~UT. The 150~MHz source was consistently imaged over a period of 8~s at approximately 0.25~R$_\odot$ and observed over the western limb, aligned with the solar equator over a large group of open magnetic field lines.

\begin{figure*}[h]
    \centering
    \includegraphics[width=0.7\textwidth]{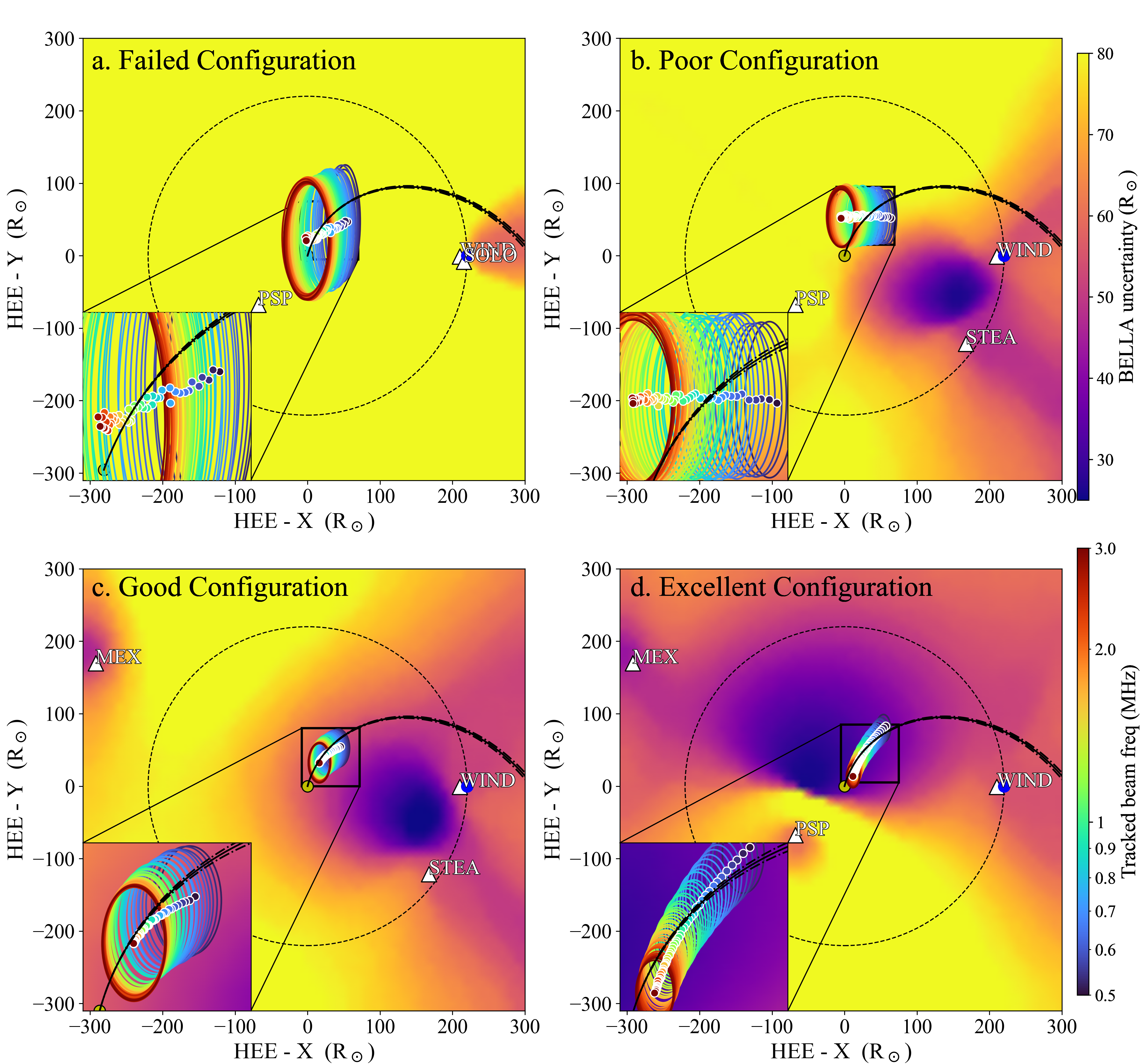}
    \caption{Radio burst source positions obtained by multilaterating different spacecraft configurations using BELLA with three spacecraft. The background maps show regions of low confidence (yellow) and high confidence (blue). Spacecraft configurations for the given event were classified as (a)~\textbf{Failed}, (b)~\textbf{Poor}, (c)~\textbf{Good}, and (d)~\textbf{Excellent}. In rainbow is shown the multilaterated Type IIIs, where the centroids are the most probable source location and the ellipses denote 1~$\sigma$ uncertainty at each corresponding frequency. The results accuracy of the multilateration are heavily dependent on the sparsity of the spacecraft. The cadence of the spacecraft will determine the lower uncertainty threshold that is accepted and therefore have an impact on the precision of the positions. A Parker Spiral was fitted to the five~spacecraft multilateration shown in Fig.~\ref{fig:5sc} and plotted over all combinations of spacecraft for reference. Assuming this Parker spiral as the ground truth, it is shown that as the spacecraft configuration becomes sparsely distributed around the burst source location, the results become more accurate and precise.}
    \label{fig:sc_config}
\end{figure*}

\begin{figure*}[h]
    \centering
    \includegraphics[width=0.7\linewidth]{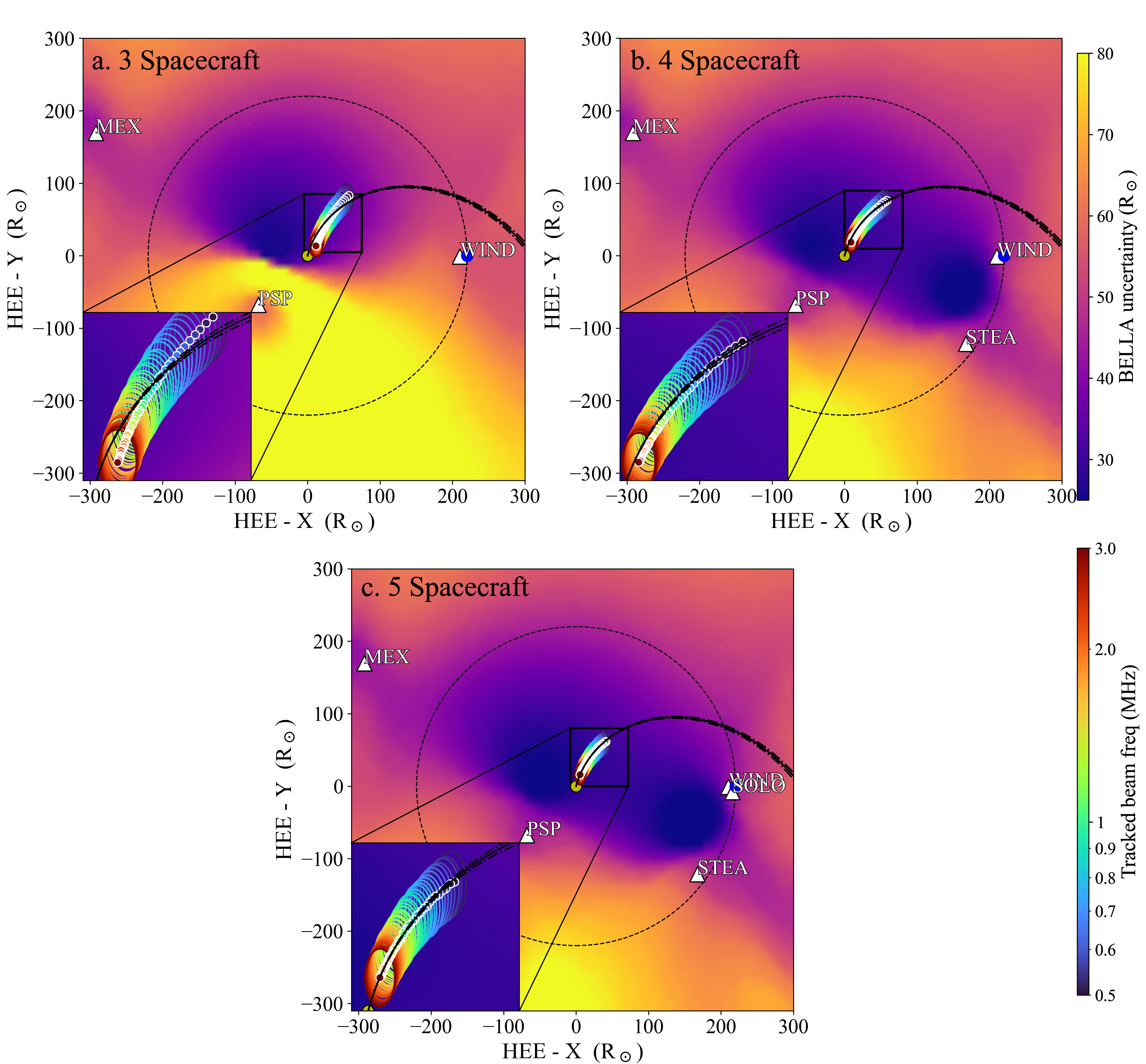}
    \caption{Multilateration of the Type III compared with the number of spacecraft being a) three spacecraft (from Fig.~\ref{fig:sc_config}d), b) four spacecraft, and c) five spacecraft (also shown in Fig.~\ref{fig:5sc}a). All spacecraft configurations were sparse and the radio sources path was located in regions of high confidence. It was found that, unlike spacecraft configuration (see Fig.~\ref{fig:sc_config}) the number of spacecraft had a relatively small impact in precision. However, there was a notable improvement in accuracy suggesting that a redundant number of spacecraft contribute towards cancelling out the errors from the data extraction procedure. }
    \label{app:sc_number}
\end{figure*}
\begin{figure*}[h]
    \centering
    \includegraphics[width=0.75\linewidth]{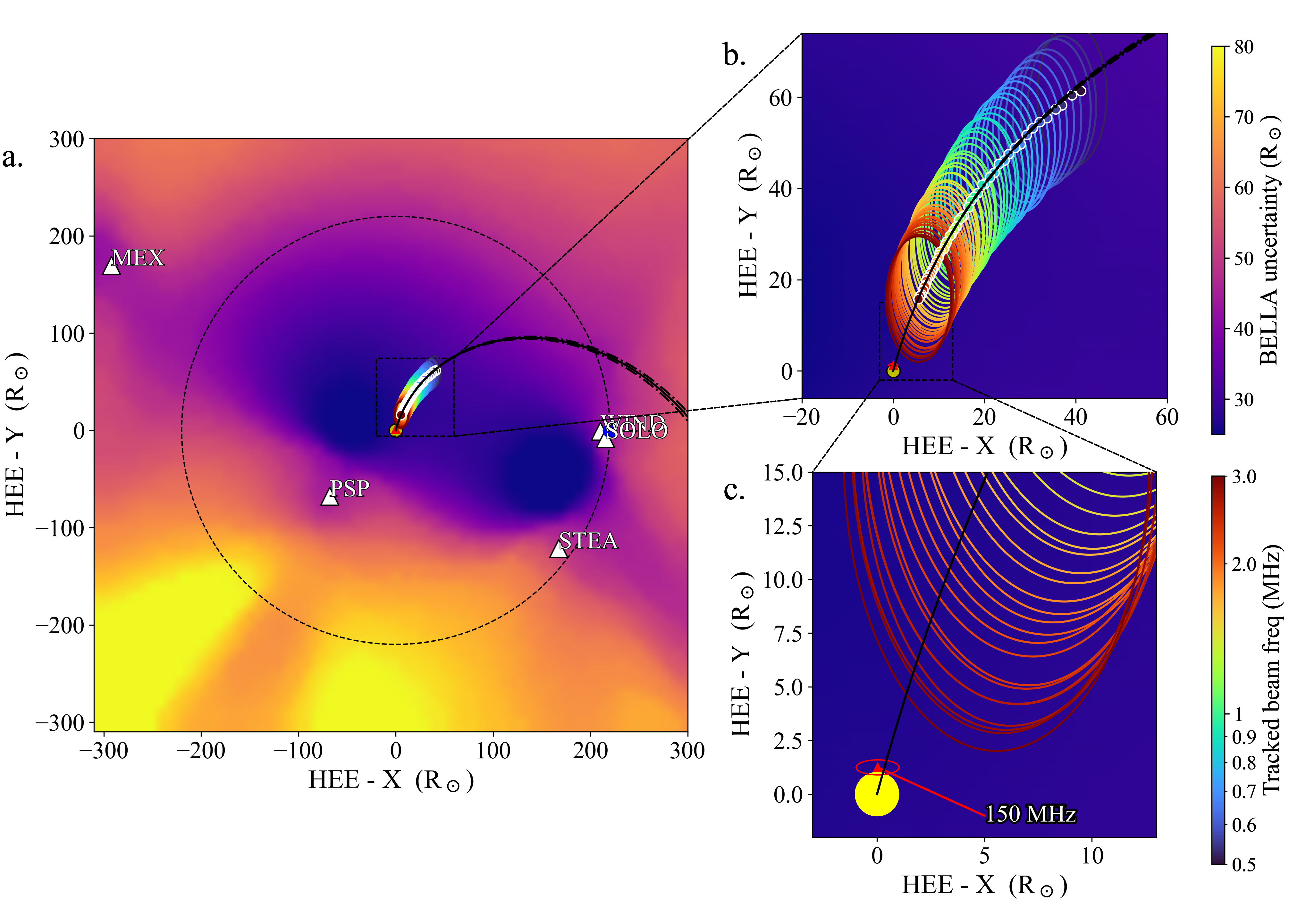}
    \caption{Multilateration of the Type~III sources using BELLA with five~spacecraft. (a)~Bird's eye view of the ecliptic plane showing the spacecraft configuration. The spacecraft configuration was relatively sparse and generated a large area of high confidence in which the BELLA sources were found. (b) Zoom of (a) showing a close up of the sources' apparent location at a frequency range of 0.5--3~MHz. A Parker Spiral was fitted to the centroids of the BELLA sources. (c) Zoom of (b) showing a clear view of the results obtained by the NRH 150~MHz interferometric imaging compared to the BELLA sources. The Parker Spiral obtained from the BELLA sources was found to be consistent with the uncertainty region of the interferometric imaging.}
    \label{fig:5sc}
\end{figure*}

At approximately the same time, GOES and STIX detected a B-class flare (see Fig.~\ref{app:xray}), however the 10--15~keV channel of STIX did not record any activity suggesting that the X-ray emission is either on-loop only or that there are not enough energetic electrons at the footpoints to produce a detectable signal. No major events were recorded an hour previous to the Type III event, therefore suggesting that the small B-class flare could be the triggering mechanism through which the electron beams got accelerated. The location of the NRH imaging will later be shown to be consistent with the apparent location of the multilaterated Type III path. Additionally, the PFSS shows open magnetic field lines forming a streamer at the western limb which, as it will be shown in Sec.~\ref{sec:propagation}, is consistent with the onset location of the open magnetic field lines of the apparent path of the electron beams suggested by the Parker Spiral derived from the multilateration results. 

\subsection{Propagation of electron beams}\label{sec:propagation}
Elementary geometry shows that the minimum number of spacecraft required for 2D multilateration is three. However, this does not mean that any three spacecraft will be capable of multilaterating a source accurately. \cite{bay:Canizares2024} used simulations that showed that the position of STEREO A and STEREO B with respect to Wind had major implications in the uncertainties of the results. In Fig.~\ref{fig:sc_config}, we empirically validate this claim through the utilisation of real observations.

Fig.~\ref{fig:dynspec} shows a Type~III radio burst detected by five different spacecraft located at different points around the heliosphere. This unique observation allows for up to 10 different combinations of three spacecraft which we use for the purposes of testing spacecraft configuration. A selection of four of the ten possible combinations of spacecraft is shown in Fig.~\ref{fig:sc_config}. In order to obtain the location of the Type~III sources and assess the uncertainty and confidence of the observation we used BELLA. The background map is obtained by simulating electromagnetic emission at every pixel of the spatial domain and the electron beams are localised using the BELLA multilateration procedure (see Sec.~\ref{sec:methods:bella}) with the front fitting functions obtained from the leading edge of the Type~IIIs shown in Fig.~\ref{fig:dynspec}. The background or confidence maps serve to identify where in the spatial domain is the multilateration method capable of obtaining results with the least possible uncertainty. This means that if a source is localised in the regions of low confidence, shown as yellow in Fig.~\ref{fig:sc_config}, then it is advisable to proceed with caution or discard the event for the purposes of multilateration. However if the source locations are localised within a high confidence area, shown in blue in Fig.~\ref{fig:sc_config}, then one can assume that the contribution of BELLA to the uncertainty of the results is negligible and any source of uncertainty is due to other factors such as physical uncertainties or instrumental. 

In order to categorise a spacecraft configuration for a given event, we define the uncertainty tolerances. The lower limit is dictated by the instrumental cadence uncertainty: 
\begin{equation}\label{eq:uncert_low}
    \Delta d_l = \frac{c \ \delta t}{r_{\odot}}
\end{equation}
where $\Delta d$ is the uncertainty limit and $\delta t$ is the cadence of the instrument. The constant $r_{\odot}$ corresponds to the value of a solar radius in metres and  converts the equation to $R_{\odot}$ units. This corresponds to $\sim$~3\,$R_\odot$ for a cadence of 7\,s and $\sim$~25\,$R_\odot$ for a cadence of 60\,s. As a rule of thumb we take the conservative approach and make confidence maps with the worst available cadence, that is 60\,s. Therefore, the lower limit is established as $\Delta d_l = 25\,R_\odot$ and defines areas of high confidence. The upper tolerance limit is selected as an arbitrary conservative constant for which any larger of an uncertainty is deemed unacceptable. For this case $\Delta d_h = 80\,R_{\odot}$. Having set the uncertainty tolerances, we categorise spacecraft configuration in the following matter:

\begin{itemize}
    \item \textbf{Failed}: Ill-defined configurations or configurations that are mostly dominated by areas with uncertainties above the upper tolerance limit. 
    \item \textbf{Poor}: Areas of confidence are present but the region is dominated by areas of low confidence. Radio sources likely to appear on a region of low confidence. 
    \item \textbf{Good}: Areas of confidence are present, areas of moderate confidence dominate the region around the Sun. 
    \item \textbf{Excellent}: Areas of confidence dominate the region around the Sun, the radio sources will likely be localised in a region of high confidence. 
\end{itemize}

Fig.~\ref{fig:sc_config} shows the multilateration results obtained by BELLA for four different configurations of three spacecraft. The Type~III was multilaterated in all cases in the range of 3--0.5~MHz and were arranged by order of confidence. Fig.~\ref{fig:sc_config}a shows an example of the \textbf{Failed} spacecraft configuration where all spacecraft are in a line and Wind and SolO are at approximately the same location. As a result, the uncertainties observed are in the order of $\sim$65~R$_\odot$ in the Heliocentric Earth ecliptic (HEE)-X direction and $\sim$160~R$_\odot$ in the HEE-Y direction showing an improvement in the uncertainties of the HEE-X direction thanks to the separation between PSP and the SolO/Wind pair. Fig.~\ref{fig:sc_config}b shows an example of a \textbf{poor} spacecraft configuration. In this case the spacecraft are sparsely distributed but the multilaterated sources are located outside of the region formed by the spacecraft. This results in uncertainties of the order $\sim$40~R$_\odot$ in HEE-X and $\sim$80~R$_\odot$ in HEE-Y. The third case shown in Fig.~\ref{fig:sc_config}c is an example of a \textbf{Good} combination of spacecraft where the spacecraft are distributed around the sources but the separation is not sparse. In this case the beam path is located in an area of moderate confidence showing uncertainties of approximately $\sim$30~R$_\odot$ in HEE-X and $\sim$60~R$_\odot$ in HEE-Y. The last case study, Fig.~\ref{fig:sc_config}d, is an example of an \textbf{Excellent} spacecraft configuration where the spacecraft positions are well separated and well distributed around the Sun, and as a consequence, the sources are located in the region of confidence. In this case scenario we see that the sources are located within an uncertainty of $\sim$18~R$_\odot$ in HEE-X and $\sim$31~R$_\odot$ in HEE-Y which is a level of precision comparable to the lowest spacecraft cadence (i.e., 60~s corresponds to 25~R$_\odot$). 

Throughout these four case studies, we observe that the apparent direction of the sources shifts. This means that these results show that the spacecraft configuration can have a major impact on the precision and accuracy of the multilateration results, which is consistent with simulation results of \cite{bay:Canizares2024}. It is therefore necessary to always use background confidence maps such as the ones computed by BELLA, for the purpose of understanding if the results obtained are physical or corrupted by the localisation method.  

Fortunately, this five spacecraft observation also allows to compare the impact of spacecraft number to the multilateration results. Fig.~\ref{app:sc_number}, shows a comparison of three, four and five spacecraft, showing a small but noticeable improvement in precision ($\sim$15~R$_\odot$ in HEE-X and $\sim$28~R$_\odot$ in HEE-Y  for the five spacecraft case) and a notable improvement in accuracy assuming a Parker Spiral as the ground truth. Comparing the three, four and five spacecraft multilateration, one can notice a change in the angle of the sources path showing that with an increase in spacecraft number the source's path aligns itself with the Sun's Parker Spiral. This apparent rotation is a consequence of extracting data from a relatively low cadence instrument. BELLA being a statistical method, uses the redundant number of spacecraft to cancel out any errors derived from the data extraction. This was addressed by concept missions such as SURROUND (\textcolor{blue}{Weigt and Cañizares et al,} \citeyear{bay:weigtcanizares2023}) which highlight the importance of multi-spacecraft missions for the localisation accuracy and precision.

\begin{figure}[h]
    \centering
    \includegraphics[width=1\linewidth]{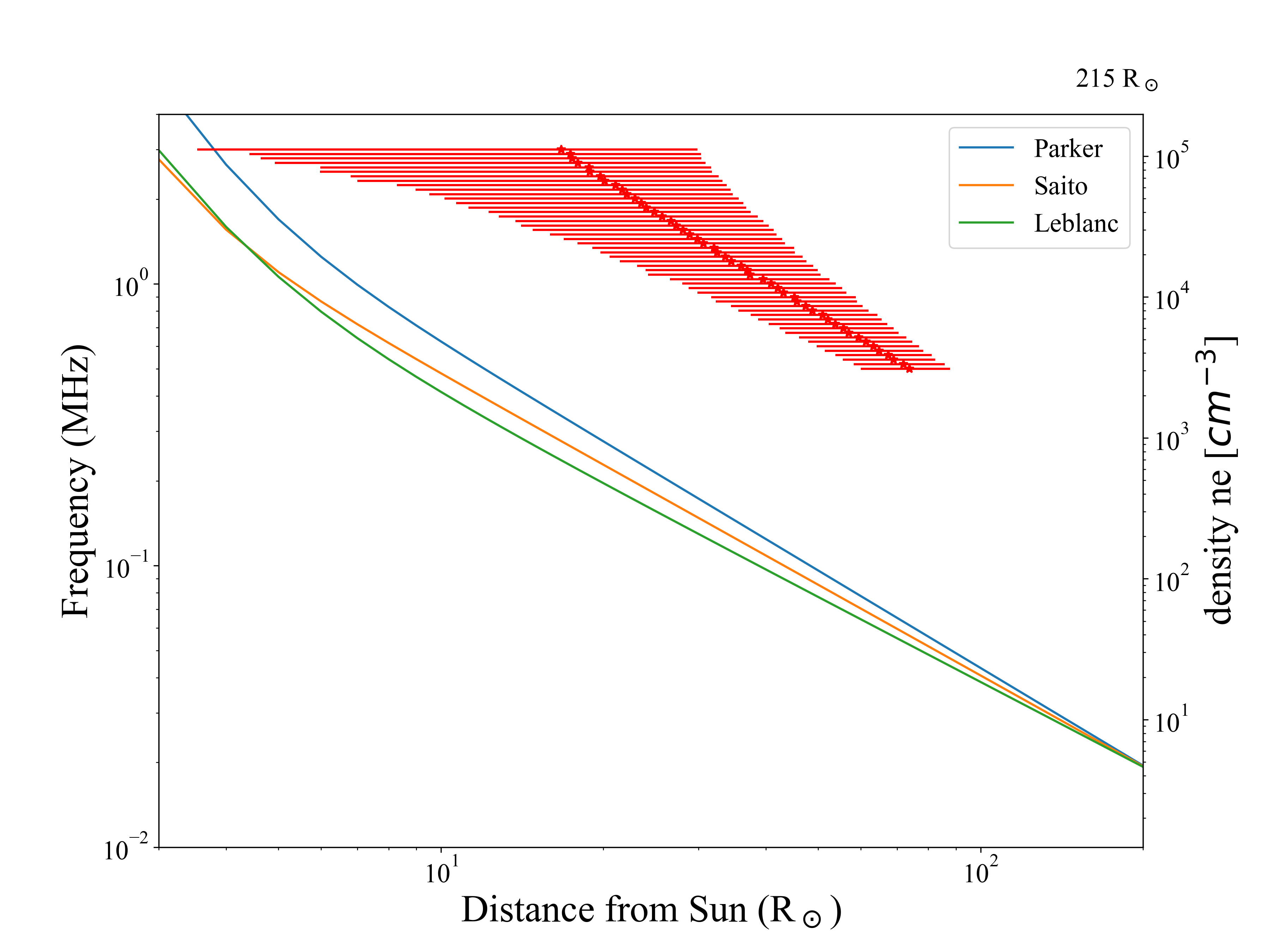}
    \caption{Different density models compared to the results obtained by BELLA. The density models were calibrated using ACE in-situ data at 1~au (see Fig.~\ref{app:ace} \textit{left}). The electron density trend shows an approximate negative $1/R^2$ slope consistent with the slope of interplanetary density models. However, the density models showed to be one order of magnitude closer to the Sun than those measured by BELLA. This is consistent with studies \citep{scat:chrysaphi2018cme,tri:chen2023source} that showed that radio scattering makes the apparent source location appear further from the Sun than their true location.}
    \label{app:densities}
\end{figure}

Fig.~\ref{fig:5sc} shows the results of performing multilateration with five spacecraft. Fig.~\ref{fig:5sc}a shows a bird's eye view of the ecliptic, showing the \textbf{Excellent} spacecraft configuration, with a relatively large area of confidence. Unfortunately, there was no spacecraft in the +/+ quadrant and therefore the Y-coordinate shows a larger uncertainty than the X-coordinate. However, the beams path is located in the large area of confidence and therefore the lack of a spacecraft in the +/+ quadrant did not have a significant impact in the multilateration. Fig.~\ref{fig:5sc}b shows a zoomed in view of Fig.~\ref{fig:5sc}a showing the localised sources. A Parker Spiral was fitted to the sources as it will be shown in Sec.~\ref{sec:scattering}. This Parker Spiral was included in all cases of Fig.~\ref{fig:sc_config} and Fig.~\ref{app:sc_number}, serving as a reference to assess the results accuracy. Fig.~\ref{fig:5sc}c shows a zoomed in view of Fig.~\ref{fig:5sc}b, showing the deprojected location of the NRH 150~MHz interferometric imaging shown in Fig.~\ref{fig:imaging}. Similarly to \cite{tri:badman2022tracking}, the deprojection was obtained by constraining the angles derived from the interferometric imaging with a Newkirk density model \citep{dm:newkirk1961solar} at one to four fold. This resulted in a series of intercepts, which are collected in the area of uncertainty. The average value of all these intersections is shown as the centroid. We observed that the 150~MHz location is consistent with the Parker Spiral derived from the BELLA sources. The agreement between the interferometric imaging and the Parker Spiral derived from the BELLA multilateration served as a validation mechanism for the confidence of location of the BELLA sources. 

\begin{figure*}[h]
    \centering
    \includegraphics[width=0.8\linewidth]{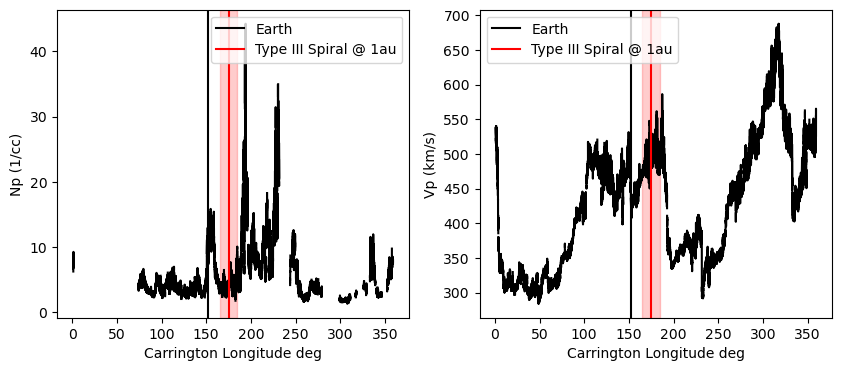}
    \caption{In situ data obtained by the ACE spacecraft. The \textit{left} plot shows the proton density and the \textit{right} plot shows the measured velocity of the solar wind. Black vertical lines corresponds to the location of the spacecraft along the Carrington longitude on the date of the event. The red vertical line corresponds to the measurements made when the spacecraft is at the Carrington longitude determined by the BELLA derived Parker Spiral at 1~au. The red shaded region corresponds to a $\pm$~10$^\circ$ around the red line to account for in-situ measurements variability. Our measurements show a the proton density of 4$\pm$2.3 1/cc and a velocity of the solar wind of 500$\pm$70~km~s$^{-1}$}
    \label{app:ace}
\end{figure*}
\begin{figure*}[h]
    \centering
    \includegraphics[width=0.8\linewidth]{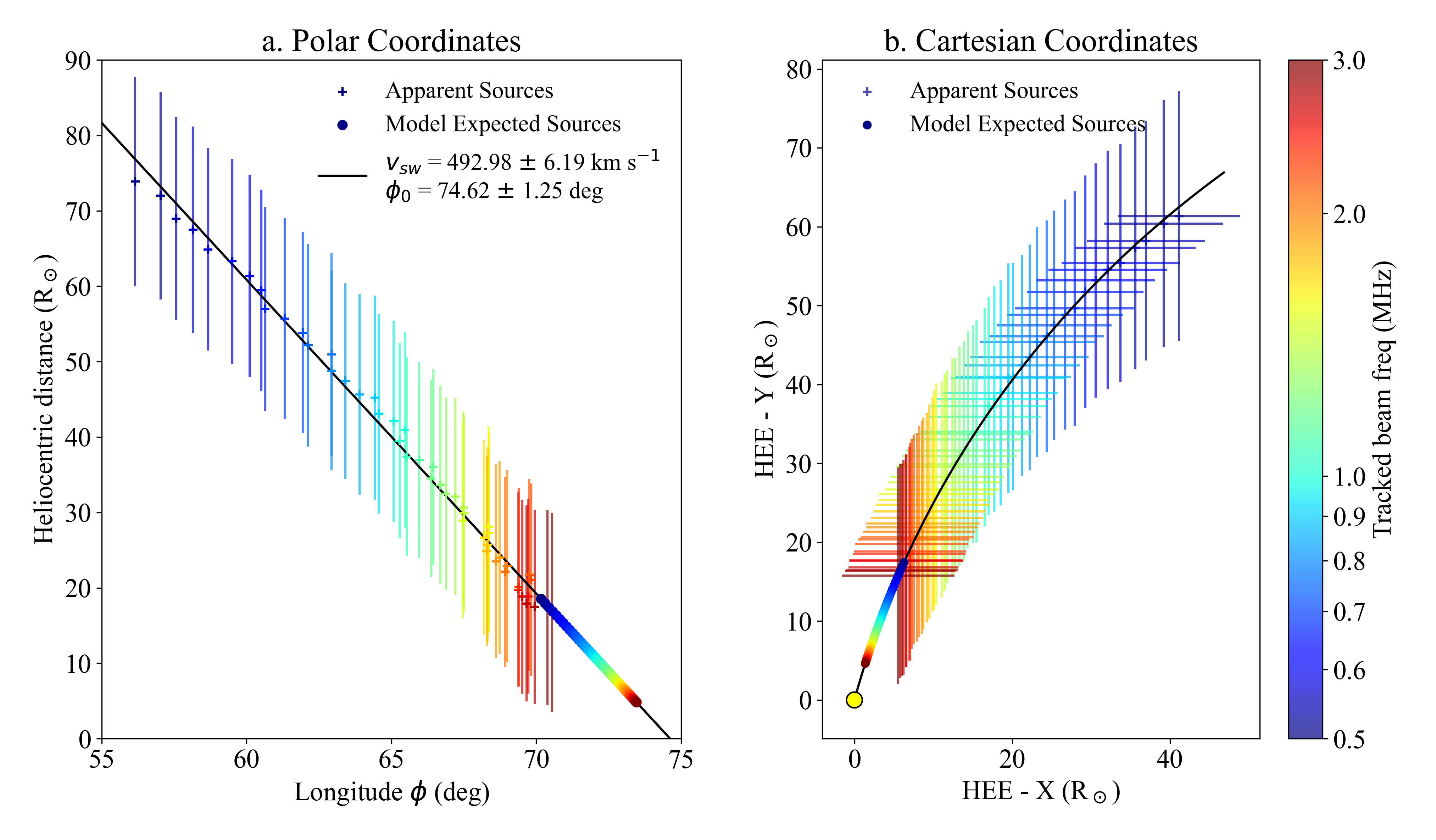}
    \caption{The positions of the apparent sources localised by BELLA compared to the expected scattering-free positions determined by the Parker density model. The localisations are shown in (a) in polar coordinates and (b) cartesian coordinates  for the readers' convenience. An Archimedean Parker spiral is a linear function in polar coordinates, the linear trend in (b) shows that the BELLA localised sources are consistent with the path of a Parker Spiral. The fitted spiral results in a velocity of the solar wind of $V_{sw}=492.98\pm6.19$~km~s$^{-1}$  and a source longitude of $\phi_0 = 74.62^\circ \pm 1.25 ^\circ$.}
    \label{fig:rphi}
\end{figure*}

\begin{figure}[h!]
    \centering
    \includegraphics[width=1\linewidth]{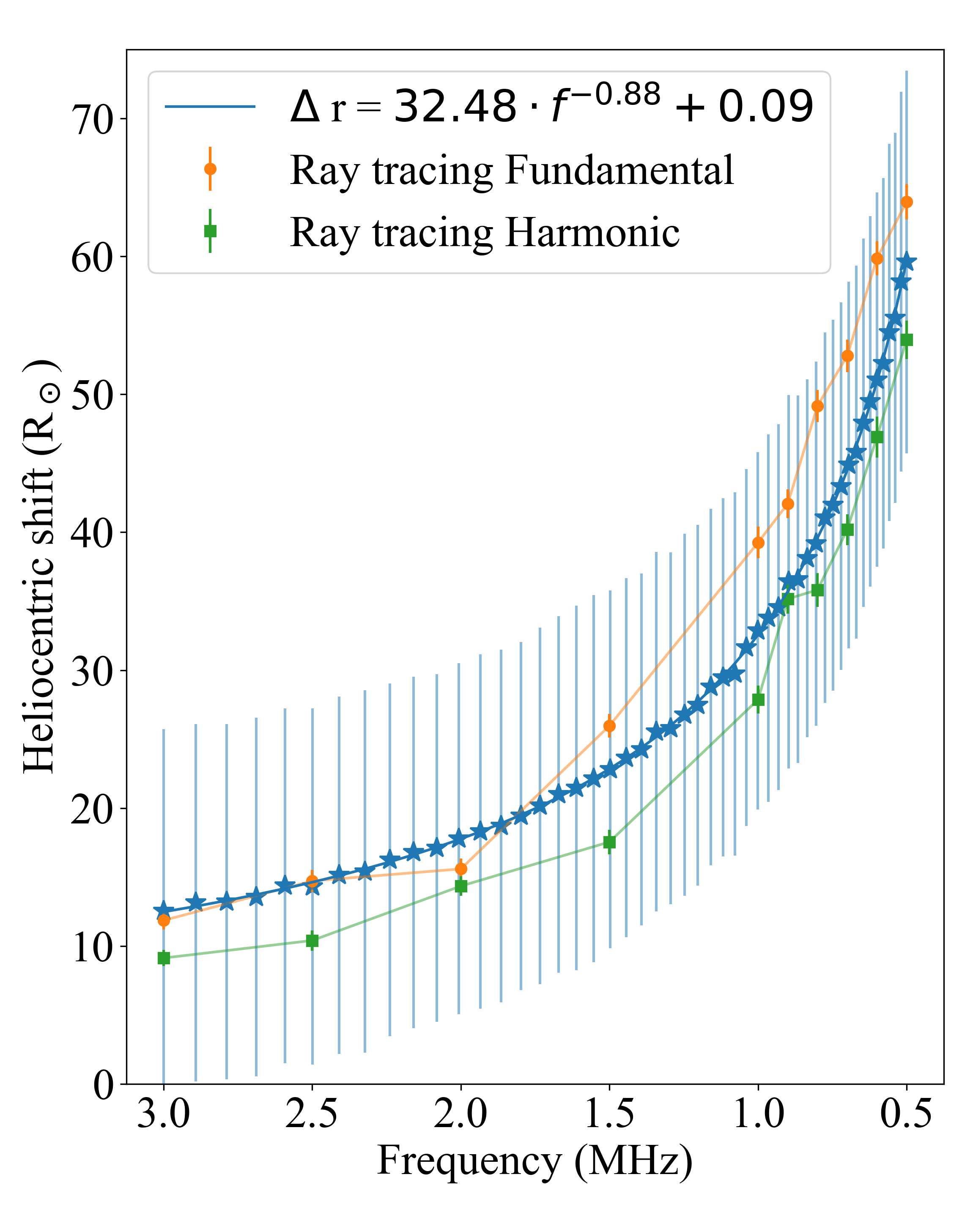}
    \caption{Scattering shifts estimated by comparing the BELLA positions to the expected positions along the Parker Spiral given by the Parker density model. Note the frequency axis is inverted as a proxy for distance from the Sun. The results from ray-tracing simulations \citep{scat:kontar2019anisotropic} are shown as orange dots (for fundamental emissions) and green squares (for harmonic emissions). BELLA was found to be in significant agreement with the ray-tracing simulations regardless of whether the emission is fundamental or harmonic. The agreement between the BELLA estimates and the ray-tracing simulations is evidence of the powerful effects that radio-wave scattering induces on the apparent position of radio sources, and validates the descriptions of the anisotropic nature of scattering in the solar corona. }
    \label{fig:scatteringshifts}
\end{figure}

\subsection{Estimating radio-wave scattering effects}\label{sec:scattering}
The results obtained by the five-spacecraft configuration (see Fig.~\ref{fig:5sc}) were the most confident case, therefore we took these measurements as the base for our analysis. Fig.~\ref{app:densities} shows the electron densities and frequencies of the emission as a function of the distance from the Sun. In this plot, we have compared the BELLA measurements with the Parker \citep{dm:parker1958dynamics}, Saito \citep{dm:saito1977} and Leblanc \citep{dm:Leblanc1998} density models which were calibrated using in-situ proton densities (assuming quasi-neutrality) measured by the Advanced Composition Explorer \citep[ACE;][]{inst:stone1998advanced} obtained after the fact, when the spacecraft was located at the corresponding Carrington longitude (see Fig.~\ref{app:ace} \textit{left}). Comparing the BELLA localisations with the density models showed higher than expected electron densities. This discrepancy was expected and is consistent with previous studies where radio-wave scattering is attributed as the dominant cause for the apparent shift and higher densities \citep{scat:chrysaphi2018cme}. 

Fig.~\ref{fig:rphi} shows the positions of the BELLA sources in (a) cartesian coordinates and (b) polar coordinates. As shown in \cite{bay:Canizares2024} we make the assumption that the Parker Spiral is Archimedean, which is linear in polar coordinates and the velocity of the solar wind is obtained from the slope and the spiral source longitude $\phi_0$ is evaluated at the intercept $r=0$:
\begin{equation}
   \phi(r) = \frac{\Omega}{v_{sw}} (r - r_0) + \phi_0
\end{equation}
Taking the centroids of the ellipses to fit an Archimedean Parker Spiral results in a $v_{sw} = 492.98 \pm 6.19$~km~s$^{-1}$ and $\phi_0 = 74.62 \pm 1.25 $~deg with a correlation coefficient $\rho^2$ = 0.992. These results were later compared with the solar wind velocity measured by the ACE spacecraft in-situ at the corresponding Carrington longitude. ACE's measurements of the solar wind (see Fig.~\ref{app:ace} \textit{right}) were found to be approximately 500$\pm$70~km~s$^{-1}$, in excellent agreement with our BELLA estimation.

Studies have shown \citep{scat:kontar2019anisotropic, scat:Kuznetsov2020, tri:chen2023source} that the density inhomogeneities are distributed anisotropically and aligned along the magnetic field line. This means that in order to correct for scattering, we start from the assumption that the sources are shifted along the BELLA derived Parker Spiral. This assumption was supported by the location of the NRH interferometric imaging and the consistency of the inferred spiral with the 1 au solar wind speed. Therefore, by using a Parker model along the spiral we obtain the theoretical location of the sources. Fig.~\ref{fig:rphi} shows the theoretical source location as dots, which can be compared to the BELLA localised sources shown as crosses. It is evident that a very large discrepancy exists between the model-expected source locations and the BELLA-estimated (observed) locations, in line with previous quantitative estimations of the scattering-induced source shifts \citep{scat:chrysaphi2018cme, tri:chen2023source}.
Therefore, a comparison of the model-expected and BELLA-estimated locations provides an approximate quantification of the scattering-induced shift that impacts the analysed radio emissions.


Fig.~\ref{fig:scatteringshifts} shows the estimated shifts as a function of frequency in the radial direction. These shifts were quantised by subtracting the heliocentric location of the theoretical `expected' sources and the heliocentric coordinate of the BELLA sources. These heliocentric shifts showed, as it was anticipated from theory, that scattering is more pronounced at the lower frequencies than at higher frequencies. By fitting a power law to these shifts, we analyse the nature of the estimated scattering. The power law was found to be:
\begin{equation}
    \Delta r = 32.48 \cdot f^{-0.88} + 0.09
\end{equation}
This power law shows that there is a large amplitude coefficient, suggesting a strong sensitivity to changes in frequency. The negative power law coefficient indicates, as expected, an inverse relationship between scattering shifts and frequency. Additionally the power law coefficient is close to -1 which indicates a nearly inversely proportional relationship between the two. It is worth noting that this $\sim f^{-1}$ relationship is also found in measurements of the source size \citep{scat:kontar2019anisotropic}, the decay time \citep{vecchio_accepted}, and the rise time \citep{scat:chrysaphi2024first}. Furthermore, the offset coefficient of the power law is close to 0, implying that as the frequency approaches infinity, the scattering shift approaches a negligible asymptotic limit, which is also expected from theory. 

Fig.~\ref{fig:scatteringshifts} also shows the expected scattering-induced shifts obtained from 3D ray-tracing simulations of radio-wave propagation in a medium of anisotropic density fluctuations, presented in \cite{scat:kontar2019anisotropic}. Simulations were performed for both fundamental and harmonic emissions, taking the Parker Spiral as the configuration of the magnetic field. The parameters of these simulations are described in Sec.~\ref{sec:methods:scatteringsimulations}.

The ray-tracing simulation outputs are in agreement to the estimates obtained by BELLA. The fundamental ray-tracing simulations are particularly in close agreement with the BELLA estimates at the higher frequency range, whereas the lower-frequency observations are between the simulated fundamental and harmonic source locations. Nevertheless, both the fundamental and harmonic simulations are consistent with the observations and their uncertainties.

The significant agreement between the two independent methods, BELLA and ray-tracing simulations, serves as confirmation of the recent advancements in our theoretical understanding and modelling of anisotropic radio-wave propagation effects \citep[e.g.][]{scat:kontar2019anisotropic, scat:Kuznetsov2020, tri:chen2023source}. Moreover, it also suggests that density inhomogeneities, which are aligned with the Parker Spiral, have a very strong effect on the path radio-burst photons take from emission to receipt.

\section{Conclusions}\label{sec:conclusions}
In this study, we presented a detailed analysis of a Type~III solar radio burst tracked from the low solar corona to interplanetary distances. By using a combination of ground-based observations from the NRH and multi-spacecraft radio data from PSP, SolO, STEREO A, Wind, and MEX we have obtained the apparent path of the Type III drivers with relatively high confidence. By using a five spacecraft observation we have achieved a precision of $\sim$~15--28~R$_\odot$ in the localisation of the apparent Type III sources, observed to fit a Parker spiral with a 99.2\% fit quality. 

The methods of positioning used in this study were interferometric imaging, obtained by NRH at 150~MHz, and BELLA at 3--0.5~MHz. We found both methods to be in agreement with each other showing that the Parker Spiral obtained by BELLA extrapolated through the uncertainty region of the interferometric imaging.   

We also showed the critical importance of spacecraft configuration in accurately localising the position of Type~III radio sources. The inclusion of MEX/MARSIS data was particularly significant for this study, demonstrating that a well-distributed and well-spaced spacecraft configuration can significantly enhance localisation accuracy and reduce uncertainties. We highlight the necessity of using confidence maps in multilateration to ensure the reliability of the localisation results as we have shown that the wrong spacecraft configuration can corrupt the apparent location of the sources.  

Furthermore, we measured the impact of radio-wave scattering on the observed positions of Type~III sources. Our measurements indicate that scattering caused a significant shift in the apparent positions of the sources along the Parker spiral. By comparing our observed positions with density models, we made estimates of the scattering shifts, revealing a strong inverse relationship between frequency and scattering shift, which was found to be in significant agreement with independent scattering simulations. This agreement adds to the existing evidence that field-aligned anisotropic density turbulence has a very strong effect on the path radio-burst photons take from emission to receipt. The close agreement between the ray-tracing simulations and the BELLA estimates, along with the consistency of the inferred Parker spiral and 1 au in-situ measurements, show that the apparent radio sources preserve physical information about the trajectory along which the electrons travel.  This opens the door to detailed single event measurements in which the apparent source trajectory can be used to iterate simulation input parameters and therefore infer physical properties of the electron beam and solar wind over interplanetary distances.

\begin{acknowledgements}
L.A.C. and the research conducted in this publication were supported by the Irish Research Council under grant number GOIPG/2019/2843. 
BELLA is available at \url{https://github.com/TCDSolar/BELLA_Multilateration} 

N.C. acknowledges funding support from the Initiative Physique des Infinis (IPI), a research training program of the Idex SUPER at Sorbonne Universit\'{e}.

B.S.-C. acknowledges support through STFC Ernest Rutherford Fellowship ST/V004115/1.

The SDO/AIA images were obtained by courtesy of Lockheed Martin Solar and Astrophysics Laboratory (LMSAL) SOLARSOFT \url{https://www.lmsal.com/solarsoft/latest_events/} 

The radio-wave propagation simulations used in this study can be found at \url{https://github.com/edkontar/radio_waves}.

\end{acknowledgements}

\section{Data availability}
The WAVES dataset was obtained from \cite{data:WindL2}, the SWAVES data was obtained from \cite{data:L3STEREOGP}, the RPW and STIX data was obtained from the Solar Orbiter Archive (SOAR) available at \url{https://soar.esac.esa.int/soar/}, the FIELDS data is available at \url{https://fields.ssl.berkeley.edu/data/} and the MARSIS data is available at \url{https://archives.esac.esa.int/psa/ftp/MARS-EXPRESS/MARSIS/}. The NRH data is available at \url{https://rsdb.obs-nancay.fr/}. The GOES/XRS data was provided by the National Oceanic and Atmospheric Administration (NOAA)'s Space Weather Prediction Center (SWPC) \url{https://services.swpc.noaa.gov/json/goes/}.

\bibliography{references.bib}{}
\bibliographystyle{aasjournal}

\appendix
\section{Extended Data}\label{secA1}
\begin{figure}[h]
    \centering
    \includegraphics[width=1\linewidth]{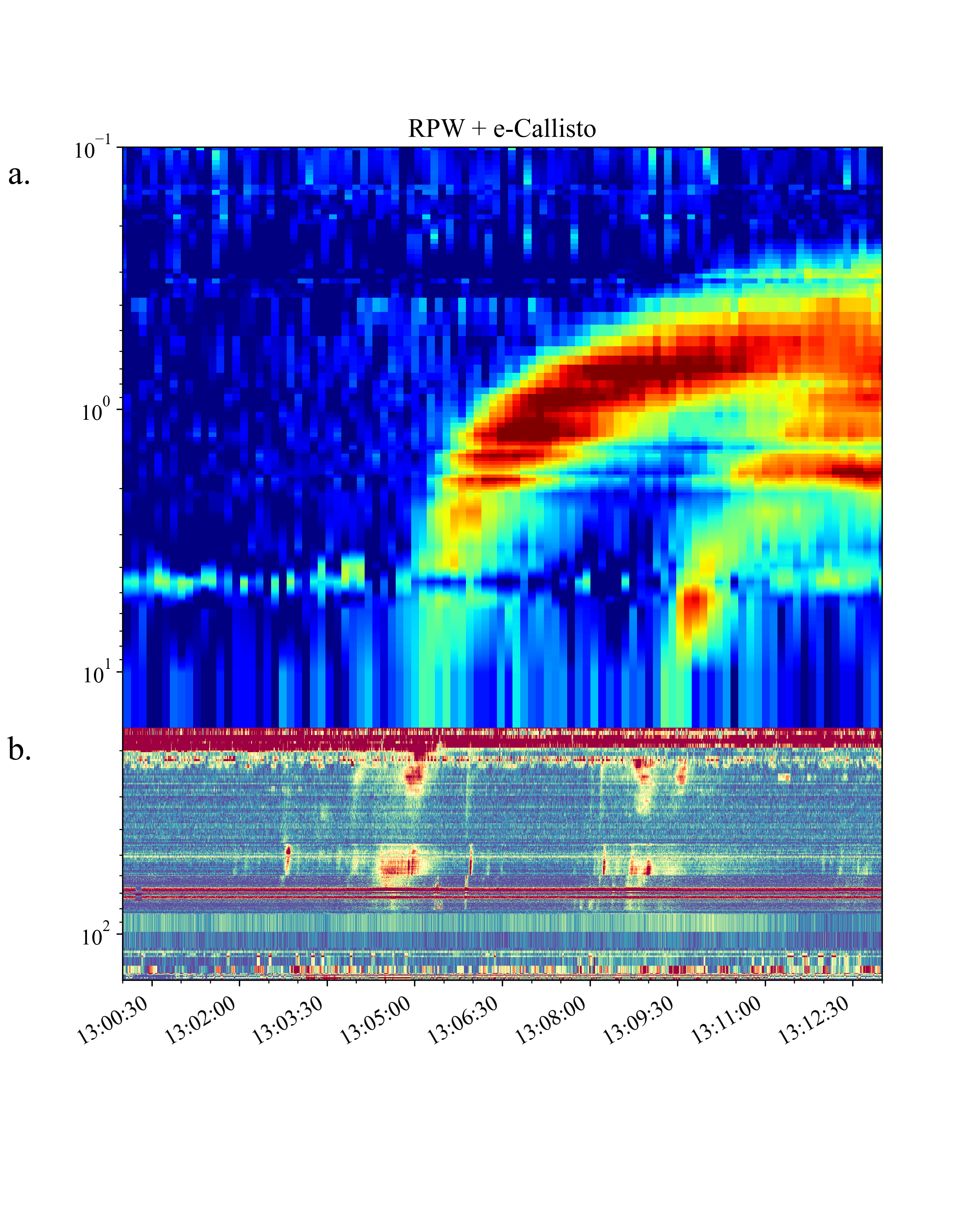}
    \caption{Combined dynamic spectra of (a) SolO/RPW and (b) e-Callisto Birr, Humain and Glasgow showing the burst radio emission at 20-80~MHz frequencies. Radio interference obscured the 150~MHz frequency channels. NRH imaging at 13:04:17.98 shows the Type III radio source which is believed to be too faint to be detected in the ORFEES spectra.}
    \label{app:ecallisto}
\end{figure}
\begin{figure}[h]
    \centering
    \includegraphics[width=1\linewidth]{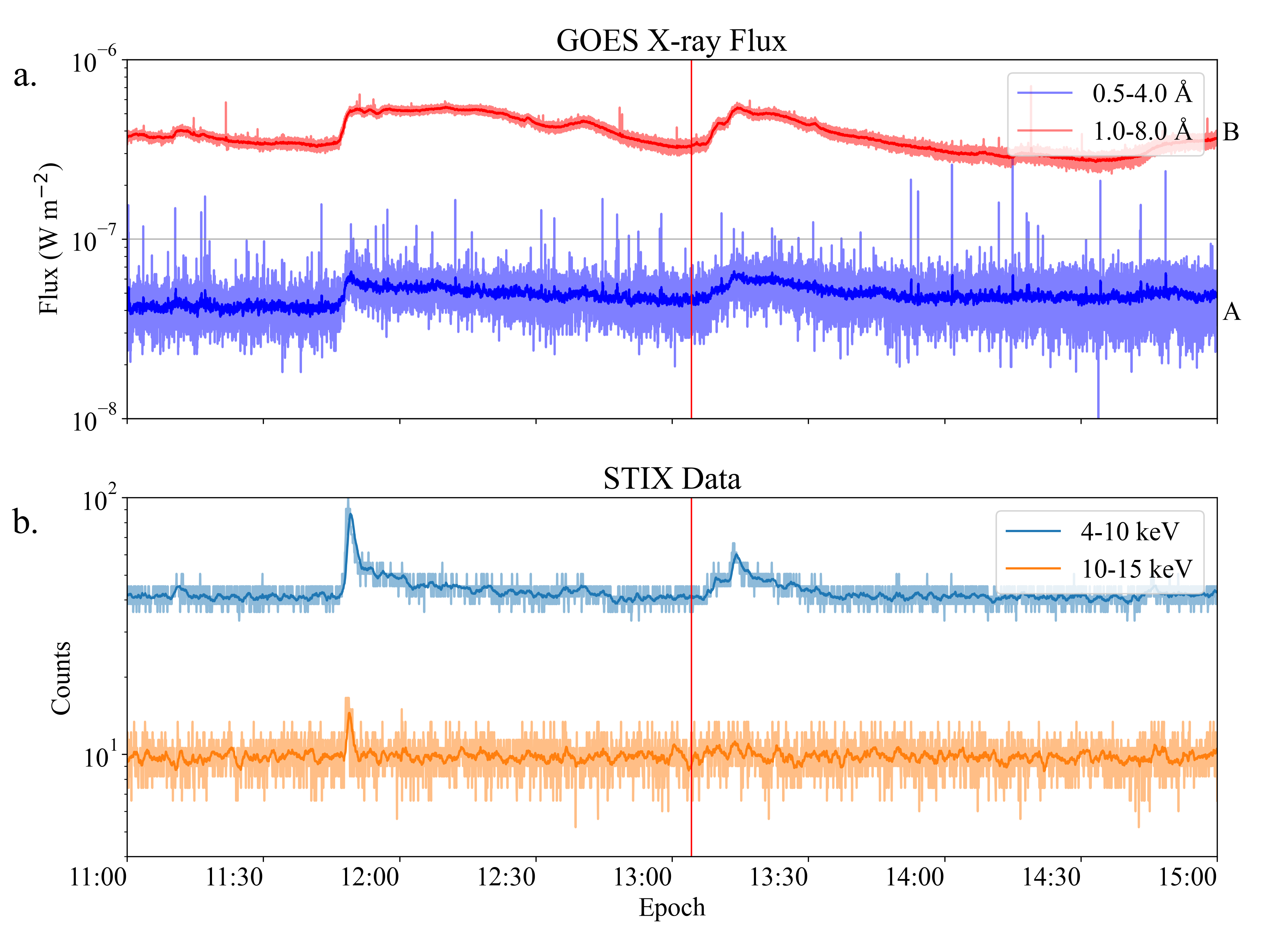}
    \caption{GOES (1.0--8.0~\AA) and SolO/STIX (4--10~keV) detection of the B class flare at $\sim$~13:04 UT that is believed to have caused the Type~III radio burst. The timing and location of the flare is consistent with the Type~III detection shown as a red vertical line.}
    \label{app:xray}
\end{figure}


    
\end{document}